\documentstyle[12pt,epsf]{article} \hoffset -0.5in \textwidth 6.00in
\textheight
8.5in  \parskip 7pt \openup1.5\jot \parindent=0.5in
\newskip\humongous \humongous=0pt plus 1000pt minus 1000pt
\def\caja{\mathsurround=0pt}
\def\eqalign#1{\,\vcenter{\openup1\jot \caja
        \ialign{\strut \hfil$\displaystyle{##}$&$
        \displaystyle{{}##}$\hfil\crcr#1\crcr}}\,}
\newif\ifdtup

\def\eqright #1\cr{\noalign{\hfill$\displaystyle{{}#1}$}}
\def\eqleft #1\cr{\noalign{\noindent$\displaystyle{{}#1}$\hfill}}

\def\oldreffmt#1{\rlap{[#1]} \hbox to 2\parindent{}}

\def\figfmt#1{\rlap{Figure {#1}} \hbox to 1in{}}

\def\sectioneq{\def\theequation{\thesection.\arabic{equation}}{\let
\holdsection=\section\def\section{\setcounter{equation}{0}\holdsection}}}%

\def\auto{\eqno(\refstepcounter{equation}\theequation)}
\def\begineq #1\endeq{$$ \refstepcounter{equation}\eqalign{#1}\eqno
	(\theequation) $$}
\def\contlimit{\,{\hbox{$\longrightarrow$}\kern-1.8em\lower1ex
\hbox{${\scriptstyle (a\rightarrow0)}$}}\,}
\def\centeron#1#2{{\setbox0=\hbox{#1}\setbox1=\hbox{#2}\ifdim
\wd1>\wd0\kern.5\wd1\kern-.5\wd0\fi
\copy0\kern-.5\wd0\kern-.5\wd1\copy1\ifdim\wd0>\wd1
\kern.5\wd0\kern-.5\wd1\fi}}
\def\centerover#1#2{\centeron{#1}{\setbox0=\hbox{#1}\setbox
1=\hbox{#2}\raise\ht0\hbox{\raise\dp1\hbox{\copy1}}}}
\def\centerunder#1#2{\centeron{#1}{\setbox0=\hbox{#1}\setbox
1=\hbox{#2}\lower\dp0\hbox{\lower\ht1\hbox{\copy1}}}}
\def\lsim{\;\centeron{\raise.35ex\hbox{$<$}}{\lower.65ex\hbox
{$\sim$}}\;}
\def\gsim{\;\centeron{\raise.35ex\hbox{$>$}}{\lower.65ex\hbox
{$\sim$}}\;}

\def\super#1{\ifmmode \hbox{\textsuper{#1}}\else\textsuper{#1}\fi}
\def\textsuper#1{\newcount\holdspacefactor\holdspacefactor=\spacefactor
$^{#1}$\spacefactor=\holdspacefactor}

\def\getcite#1,{\advance\citenumber by1
\def\getcitearg{#1}\def\lastarg{@}
\ifnum\citenumber=1
\ref{#1}\let\next=\getcite\else\ifx\getcitearg\lastarg\let\next=\relax
\else ,\ref{#1}\let\next=\getcite\fi\fi\next}

\def\pom{{\rm P\kern -0.53em\llap I\,}}
\def\spom{{\rm P\kern -0.36em\llap \small I\,}}
\def\sspom{{\rm P\kern -0.33em\llap \footnotesize I\,}}

\relax

\newskip\humongous \humongous=0pt plus 1000pt minus 1000pt
\def\caja{\mathsurround=0pt}
\def\eqalign#1{\,\vcenter{\openup1\jot \caja
        \ialign{\strut \hfil$\displaystyle{##}$&$
        \displaystyle{{}##}$\hfil\crcr#1\crcr}}\,}
\newif\ifdtup

\def\eqright #1\cr{\noalign{\hfill$\displaystyle{{}#1}$}}
\def\eqleft #1\cr{\noalign{\noindent$\displaystyle{{}#1}$\hfill}}

\def\oldreffmt#1{\rlap{[#1]} \hbox to 2\parindent{}}

\def\figfmt#1{\rlap{Figure {#1}} \hbox to 1in{}}

\def\auto{\eqno(\refstepcounter{equation}\theequation)}
\def\begineq #1\endeq{$$ \refstepcounter{equation}\eqalign{#1}\eqno
	(\theequation) $$}
\def\contlimit{\,{\hbox{$\longrightarrow$}\kern-1.8em\lower1ex
\hbox{${\scriptstyle (a\rightarrow0)}$}}\,}
\def\centeron#1#2{{\setbox0=\hbox{#1}\setbox1=\hbox{#2}\ifdim
\wd1>\wd0\kern.5\wd1\kern-.5\wd0\fi
\copy0\kern-.5\wd0\kern-.5\wd1\copy1\ifdim\wd0>\wd1
\kern.5\wd0\kern-.5\wd1\fi}}
\def\centerover#1#2{\centeron{#1}{\setbox0=\hbox{#1}\setbox
1=\hbox{#2}\raise\ht0\hbox{\raise\dp1\hbox{\copy1}}}}
\def\centerunder#1#2{\centeron{#1}{\setbox0=\hbox{#1}\setbox
1=\hbox{#2}\lower\dp0\hbox{\lower\ht1\hbox{\copy1}}}}
\def\lsim{\;\centeron{\raise.35ex\hbox{$<$}}{\lower.65ex\hbox
{$\sim$}}\;}
\def\gsim{\;\centeron{\raise.35ex\hbox{$>$}}{\lower.65ex\hbox
{$\sim$}}\;}

\def\super#1{\ifmmode \hbox{\textsuper{#1}}\else\textsuper{#1}\fi}
\def\textsuper#1{\newcount\holdspacefactor\holdspacefactor=\spacefactor
$^{#1}$\spacefactor=\holdspacefactor}

\def\getcite#1,{\advance\citenumber by1
\ifnum\citenumber=1
\ref{#1}\let\next=\getcite\else\ifx#1@\let\next=\relax
\else ,\ref{#1}\let\next=\getcite\fi\fi\next}

\def\upon #1/#2 {{\textstyle{#1\over #2}}}
\relax

\def\mainhead#1{\setcounter{equation}{0}\addtocounter{section}{1}
  \vbox{\begin{center}\large\bf #1\end{center}}\nobreak\par}
\sectioneq

\def\til#1{\centeron{\hbox{$#1$}}{\lower 2ex\hbox{$\char'176$}}}
\def\tild#1{\centeron{\hbox{$\,#1$}}{\lower 2.5ex\hbox{$\char'176$}}}
\def\sumtil{\centeron{\hbox{$\displaystyle\sum$}}{\lower
-1.5ex\hbox{$\widetilde{\phantom{xx}}$}}}

\def\pom{{\rm P\kern -0.53em\llap I\,}}
\def\spom{{\rm P\kern -0.36em\llap \small I\,}}
\def\sspom{{\rm P\kern -0.33em\llap \footnotesize I\,}}

\newcommand{\bit}{\begin{itemize}}
\newcommand{\eit}{\end{itemize}}

\newcommand{\beq}{\begin{equation}}
\newcommand{\eeq}{\end{equation}}
\newcommand{\beqa}{\begin{eqnarray}}
\newcommand{\eeqa}{\end{eqnarray}}

\begin{document}

\begin{titlepage}
\rightline{\vbox{\halign{&#\hfil\cr
&ANL-HEP-PR-95-53\cr
&IFTIP/BBSR/95-80\cr
&UF-IFT-HEP-95-20\cr}}}

\vspace{.4in}

\begin{center}

{\bf  SCALE INVARIANT $O(g^4)$ LIPATOV KERNELS}\par
{\bf AT NON-ZERO MOMENTUM TRANSFER}\footnote{Work supported by the U.S.
Department of Energy, Division of High Energy Physics, \newline Contracts
W-31-109-ENG-38 and DEFG05-86-ER-40272}
\medskip

{Claudio Corian\`{o},$^{a,b}$\footnote{
coriano@phys.ufl.edu ~$^{+}$parwani@iopb.ernet.in ~$^{\#}$arw@hep.anl.gov}
Rajesh R. Parwani$^{c+}$\ and \
Alan. R. White$^{a\#}$}

\vskip 0.6cm

\centerline{$^a$High Energy Physics Division}
\centerline{Argonne National Laboratory}
\centerline{9700 South Cass, Il 60439, USA.}
\vspace{0.5cm}

\centerline{$^b$Institute for Fundamental Theory}
\centerline{Department of Physics}
\centerline{ University of Florida at Gainesville, FL 32611, USA}
\vspace{0.5cm}

\centerline{$^c$Institute of Physics,}
\centerline{Bhubaneswar 751005, India.}

\end{center}

\begin{abstract}
We summarize recent work on the evaluation of the scale invariant
next-to-leading order Lipatov kernel, constructed via transverse momentum
diagrams. At zero momentum transfer the square of the leading-order kernel
appears together with an additional component, now identified as a new
partial-wave amplitude, having a separate, holomorphically
factorizable, spectrum. We present a simplified expression for the full
kernel at non-zero momentum transfer and give a complete analysis of its
infrared properties. We also construct a non-forward extension of the new
amplitude which is infra-red finite and satifies Ward identity constraints.
We conjecture that this new kernel has the conformal invariance properties
corresponding to the holomorphic factorization of the forward spectrum.

\end{abstract}

\end{titlepage}


\mainhead{1. A BRIEF OVERVIEW}

The Regge limit of QCD has recently undergone a considerable revival of
interest. The small-x behaviour of the parton distributions observed at
HERA, characterized by a strong rise of the gluon density, and the detection
of diffractive hard scattering events in DIS, both provide motivation for
developing a better understanding of the Regge regime of QCD. Because of the
overlap of the small-x and Regge limits, it is natural to expect that the
theoretical tools developed in the past in the analysis of Regge theory are
useful also at small-x. Properties of the ``exchanged reggeon singularities"
can be constructed from perturbation theory, by resumming the leading
$log\,\, 1/x$ and$/$or $log Q^2$ behaviour. Resummation is achieved in
various possible ways, but it is widely anticipated that the BFKL evolution
equation \cite{bfkl}, first derived more than 20 years ago,  plays a crucial
role in describing the physical properties of the leading ``Pomeron''
singularity.

The crucial ingredient in the ``construction" of the BFKL Pomeron is
the kernel of the evolution equation, its spectrum and its leading
eigenvalue. Both forward ($q=0$) and non-forward ($q\neq 0$) versions
of the lowest order ($O(g^2)$) kernel are known. Conformal partial waves
diagonalize the $O(g^2)$ equation at non zero $q$, since the equation is
invariant under special conformal transformations, and in the limit of $q\to
0$ reproduce the well known eigenfunctions, and eigenvalues of the BFKL
{\it parton} (or forward) kernel. A necessary condition for the conformal
invariance of the equation is the property of holomorphic factorization
of the eigenvalues of the parton kernel.

Most analyses of the BFKL equation involve only the $O(g^2)$ kernel
and its related properties of conformal invariance. It is, of course, important
to see how radiative corrections affect the leading order evolution. It is
expected that renormalization effects will introduce a running of the
coupling and will spoil conformal invariance. The direct
evaluation of next-to-leading-order(NLO) contributions to this equation
requires both a calculation of the correction to the Regge trajectory of the
gluon and corresponding corrections to the reggeon(s)-particle(s) transition
vertices. So far only part of this program has been completed\cite{FL}.

Both the leading-order kernel and an infrared approximation to the NLO order
kernel have been determined by a reggeon diagram technique based indirectly on
$t$-channel unitarity\cite{ker}. More recently we have shown\cite{uni,cw} how
these results can be obtained by a direct analysis of the
t-channel unitarity equations, analytically continued in the complex
$j-plane$ and expanded around nonsense poles. Also, in a recent paper,
Kirschner\cite{rk} has discussed how the same NLO kernel may emerge as an
approximation when non-leading results are obtained using the $s$-channel
multi-Regge effective action. The kernel obtained is automatically scale
invariant (there is no scale in the $t$-channel analysis) and is naturally
expressed in terms of two-dimensional transverse momentum integrals. We
should emphasize, nevertheless, that both the $t$-channel analysis\cite{cw}
and the $s$-channel formalism\cite{rk}, imply that the ambiguity of the
scale-dependence includes the overall normalization of the kernel.

Previously we have shown\cite{cw1} that, in the forward case, the new
NLO kernel splits naturally into two components. A part
proportional to the square of the $O(g^2)$ kernel and a new component that
is separately infrared safe and has an eigenvalue spectrum sharing many of
properties of the leading-order spectrum. In particular the very important
property of holomorphic factorization. From the unitarity analysis\cite{cw}
we have now shown that this new component is actually {\it a distinct
partial-wave amplitude that appears for the first time at $O(g^4)$.} It is
natural to expect that the spectral property of holomorphic factorization
will be related to the leading-order conformal invariance of this amplitude
when it is fully identified as a non-forward kernel.

In this work we are going to elaborate further on the non-forward, scale
invariant NLO kernel, by providing an explicit proof of its infrared safety
and simplifying drastically the expression given in \cite{cw1}. For this
purpose we extend to the non-forward case a method of calculation of the
various diagrams, based on the use of complex momenta, due to
Kirschner\cite{rk}. This method has been successful in reproducing the
spectrum calculated in \cite{cw1} and in separating, in the forward
direction, the new holomorphically factorizable component. We will
explicitly construct a non-forward extension of this component that has the
appropriate analytic structure and satisfies the Ward identity constraints.
We believe that this extension can indeed be identified as a new partial-wave
amplitude which at ``leading-order'' is conformally invariant.
We intend to study this issue in the near future. (Note that
since the spectrum of a conformally invariant kernel is independent of $q^2$
it is, in principle, defined uniquely by the forward spectrum.)

We will also show that the new non-forward (potentially conformally
invariant) kernel is not naturally written as a transverse momentum integral
but rather is simply expressed in the complex momenta formalism of
Kirschner. This is interesting because the unitarity formalism of \cite{cw}
actually shows that {\it the transverse momentum integral formalism is only
 necessarily applicable, as $q^2 \to 0$, and for the leading threshold
behaviour
in reggeon mass variables.} We show explicitly that extracting this threshold
behavior from the transverse momentum integral kernel is not sufficient to
give the desired non-forward extension. It is only at $q^2 = 0$ that
a transverse momentum integral gives the appropriate threshold behavior.

\mainhead{2. THE FORWARD KERNEL}

Consider first the leading-order BFKL evolution equation for parton
distributions at small-x i.e.
$$
{\partial \over \partial (ln {1 / x})}F(x,k^2) ~=~\tilde{F}(x,k^2)~+~
{1 \over (2\pi)^3}\int {d^2k' \over (k')^4} ~K(k,k') F(x,{k'}^2)
\auto\label{eve}
$$
with a  parton kernel $K(k,k')$ given (for $SU(N)$) by
$$
(Ng^2)^{-1}K(k,q)~=~\Biggl(-~\delta^2(k-k')k^6
\int {d^2p \over p^2(k-p)^2}~+~{2k^2{k'}^2 \over (k-k')^2}~\Biggr)
\auto
$$
The original Regge limit derivation included also a non-forward (i.e.$q
\neq 0$ in the following) version of this equation. Transforming to $\omega$
- space, where $\omega$ is conjugate to $ln~{1 \over x}$, the non-forward
equation takes the form
$$
\omega F(\omega,k,q-k) ~=~\tilde{F}~+~ {1 \over 16\pi^3}\int {d^2k' \over
(k')^2(k'-q)^2}~K(k,k',q) F(\omega,k',q-k')
\auto\label{ome}
$$
where the ``reggeon'' kernel $K(k,k',q)~=~K^{(2)}_{2,2}(q-k,k,k',q-k')$
contains three kinematic forms.
$$
\eqalign{ {1 \over Ng^2}&~K^{(2)}_{2,2}(k_1,k_2,k_3,k_4)~
 =~\sum \Biggl(-{1 \over
2}k_1^4J_1(k_1^2)k_2^2(16\pi^3)\delta^2(k_2-k_3)\cr
&+~{k_1^2k_3^2 \over (k_1-k_4)^2}
{}~~-~~{1 \over 2}(k_1+k_2)^2\Biggr)
\equiv~~K^{(2)}_1~+~K^{(2)}_2~+K^{(2)}_3 ~~.}
\auto\label{2,2}
$$
where
$$
J_1(k^2)~=~{1 \over 16\pi^3}\int {d^2k' \over
(k')^2(k'-k)^2}
\auto\label{j1}
$$
and the $\sum$ implies that we sum over {\it combined} permutations of both the
initial and the final state (i.e. $1 <-> 2$ combined with $3 <-> 4$).

We use transverse momentum diagrams, which we construct using the components
illustrated in Fig.~2.1.

\leavevmode
\epsffile{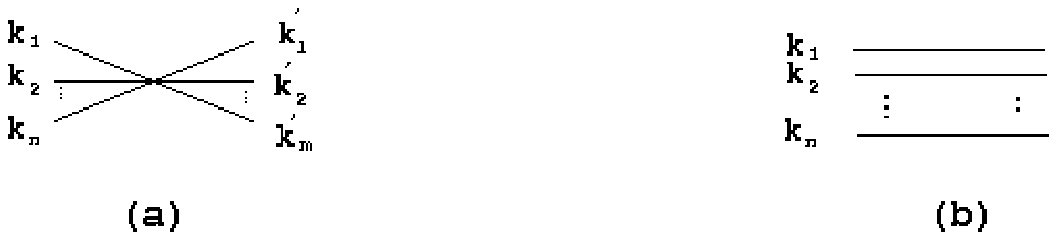}
\begin{center}

Fig.~2.1 (a)vertices and (b) intermediate states in transverse momentum.
\end{center}
The rules for writing amplitudes corresponding to the diagrams are the
following

\begin{itemize}

\item{For each vertex, illustrated in Fig.~2.1(a), we write a factor
$$
16\pi^3\delta^2(\sum k_i~  - \sum k_i')(\sum k_i~)^2
$$
}
\item{For each intermediate state, illustrated in Fig.~2.1(b), we write a
factor
$$
(16\pi^3)^{-n}\int d^2k_1...d^2k_n~ /~k_1^2...k_n^2
$$
}
\end{itemize}
Dimensionless kernels are defined by a hat
$$
\hat{K}^{(2)}_{2,2}(k_1,k_2,k_3,k_4)~=~
16\pi^3\delta^2(k_1+k_2-k_3-k_4) K^{(2)}_{2,2}(k_1,k_2,k_3,k_4)~
$$
The kernels so defined are formally
scale-invariant (even though potentially infra-red divergent). The
diagrammatic representation of $\hat{K}^{(2)}_{2,2}$, the non forward
BFKL kernel, is then as in Fig.~2.2.
\begin{center}
\leavevmode
\epsfxsize=4in
\epsffile{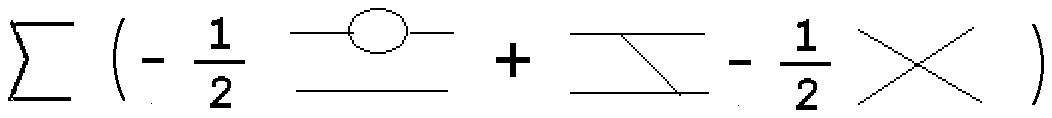}

Fig.~2.2 Diagrammatic representation of $\hat{K}^{(2)}_{2,2}$
\end{center}
The summation sign again implies a sum over combined
permutations of the initial and final momenta.

The $O(g^4)$ transverse momentum integral kernel $K^{(4)}_{2,2}$, obtained
by considering the contribution of the 4-particle nonsense states to the
unitarity equations is defined by the sum
$$
\eqalign{{1 \over (g^2N)^2} K^{(4n)}_{2,2}(k_1&,k_2,k_3,k_4)
{}~=~K^{(4)}_0~+~K^{(4)}_1~+~K^{(4)}_2~+~K^{(4)}_3~+K^{(4)}_4~}.
\auto\label{sum}
$$
with
$$
\eqalign{K^{(4)}_0~=~
\sum ~ k_1^4k_2^4J_1(k_1^2)J_1(k_2^2)(16\pi^3)\delta^2(k_2-k_3)~,}
\auto
$$

$$
\eqalign{K^{(4)}_1~=~-{2 \over 3}~
\sum ~ k_1^4J_2(k_1^2)k_2^2(16\pi^3)\delta^2(k_2-k_3)}
\auto
$$

$$
\eqalign{K^{(4)}_2~=~-  \sum
\Biggl({k_1^2J_1(k_1^2)k_2^2k_3^2+
k_1^2k_3^2J_1(k_4^2)k_4^2 \over
(k_1-k_4)^2} \Biggr),}
\auto
$$

$$
\eqalign{K^{(4)}_3~= ~\sum~
k_2^2k_4^2J_1((k_1-k_4)^2)~,}
\auto
$$
and
$$
\eqalign{K^{(4)}_4~=~{1 \over 2}~\sum~
k_1^2k_2^2k_3^2k_4^2~I(k_1,k_2,k_3,k_4), }
\auto
$$
where $J_1(k^2)$ is defined by (\ref{j1}) and
$$
\eqalign{J_2(k^2)~=~{1 \over 16\pi^3}\int d^2q {1 \over
(k-q)^2}J_1(q^2)~~~,}
\auto
$$

and
$$
\eqalign{ I(k_1,k_2,k_3,k_4)~=~{1 \over 16\pi^3}\int d^2p {1 \over
p^2(p+k_1)^2(p+k_1-k_4)^2(p+k_3)^2}.}
\auto\label{box}
$$
 The diagrammatic representation of
${\hat{K}}^{4n}_{2,2}$ is shown in Fig.~2.3.
\begin{center}
\leavevmode
\epsfxsize=5in
\epsffile{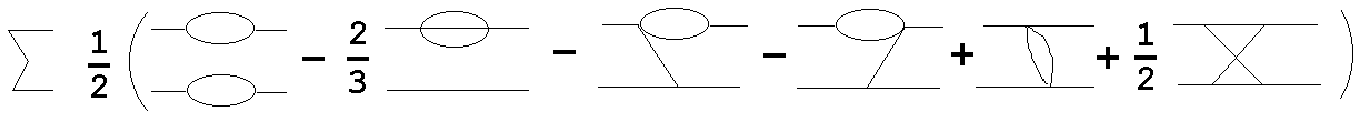}

Fig.~2.3 The diagrammatic representation of ${\hat{K}}^{4n}_{2,2}$.
\end{center}
The evaluation of these diagrams (in particular the
non planar box) has been done by an extension of
the K\"{a}llen and Toll  method \cite{KT}, developed in \cite{cw1}.
This involves a rewriting of the ``trees" \cite{KT} of the
decomposition in a suitable base. The decomposition has the advantage of
generating a minimal number of logarithms. The proliferation of
logarithms at NLO is a considerable source of complexity. (At leading order
there is a logarithm only in the trajectory function of the gluon.)
In particular, the box introduces 6 logarithms, each of which is obtained by
putting on shell 2 lines (pairwise) and which  we represent as in Fig.~2.4.
\begin{center}
\leavevmode
\epsfxsize=4in
\epsffile{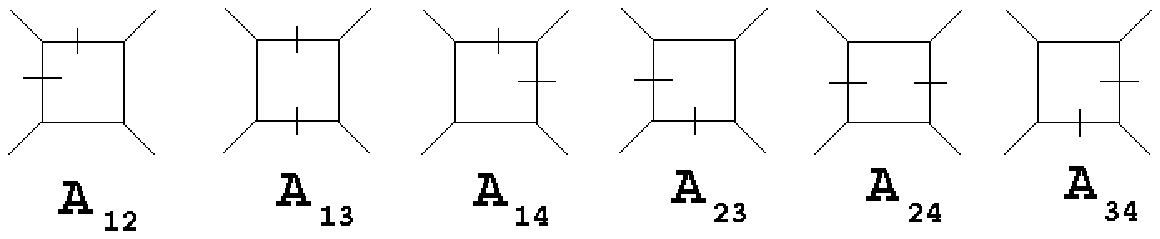}

Fig.~2.4 Tree diagrams obtained by putting on-shell the crossed lines.
\end{center}
The logarithms are of two types:

\begin{itemize}

\item[{1)}] external line ``mass'' thresholds i.e. $A_{12},A_{14},A_{23}$
and $A_{34}$  - four logarithms.

\item[{2)}]``s'' and ``t'' thresholds i.e. $A_{13},A_{24}$ - two logarithms.

\end{itemize}

In the forward direction it is straightforward to combine the type 2)
logarithms from the box with the logarithms
of the connected components $K_2^{(4)}$ and $K_3^{(4)}$
giving (in the notation of \cite{cw1} ${\cal K}_1$). Adding the
logarithms generated by  the disconnected
components $K_0^{(4)}, K_1^{(4)}$ (denoted in \cite{cw1} as ${\cal K}_0$)
gives a kernel which is infrared safe both before and after
convolution with the eigenfunctions and is equal to the $square$ of the
lowest order BFKL kernel $K_{2,2}^{(2)}$. That is we have the identity
$$
\eqalign{ \hat{{\cal K}}_0~+~ \hat{{\cal K}}_1 ~=~
{1 \over 4} \biggl(\hat{K}^{(2)}_{2,2}\biggr)^2~, }
\auto\label{k_2s}
$$
The proof of this identity is given in \cite{cw1}.

The set of box diagram logarithms 1) was denoted in \cite{cw1} as
${\cal K}_2$. It contains only the mass thresholds and is the contribution
which, in the forward direction, the unitarity analysis of \cite{cw}
determines should be correctly given by the transverse momentum integral
formalism. It is a new, separately infrared finite,
kernel for which the spectrum has been calculated and shown to satisfy the
property of holomorphic factorization\cite{cw1}. Therefore for the full
forward, or parton, kernel we can write
\beq
K_{2,2}^{(4)}= g^2 K_{BFKL} + O(g^4)\biggl[ (K_{BFKL}/2)^2 + {\cal K}_2
\biggr].
\label{complete}
\eeq
where both $K_{BFKL}$ and ${\cal K}_2$ have a spectrum which is holomorphically
factorizable. In both cases the spectrum is also infrared(IR) safe. (In
writing ``$O(g^4)$'' in (\ref{complete}) we have indicated the normalization
uncertainty due to scale dependence.)

While a direct check of IR safety is easily accomplished
in the case of the forward kernel, the case of the non forward kernel, starting
from its explicit expression given in \cite{cw1}, is far less obvious.
The proof of infrared safety given there involves diagrammatic identities.
In the next sections  we are going to reproduce this cancelation by
defining suitable, consistent, regularizations of the various components of
the kernel. We will work directly in two dimensions and show from the final
expression that the resulting kernel is IR finite. The method employs an
analytic continuation of the diagrams to complex space.

\mainhead{3. THE NON-FORWARD TRANSVERSE MOMENTUM DIAGRAM KERNEL}

Kirschner has recently shown \cite{rk} that the same separation of the
${\cal K}_2$ component from the remaining part of the $g^4$ kernel, first
obtained in \cite{cw1}, can be reobtained by performing a complex expansion
of the relevant diagrams. Here we extend his method of calculation to the
non-forward case. We complexify the ``propagators'' and ``vertices'' as
follows. We write
\beqa
 {1\over k^2} &&\to {1\over k k^*} \equiv ~~ {1\over |k|^2}\nonumber \\
 (k+q)^2(k-q)^2 &&\to |k +q|^2|k-q|^2 =~~|k^2-q^2|^2 |k'^2-q^2|^2.
\eeqa
That is we replace all the momenta $k=(k_0,k_1)$
by their complex versions $k=k_0 + i k_1$.
We also define
\beq
 R R'\equiv |k^2-q^2|^2|k'^2-q|^2
\eeq

The contribution of the box diagram to $K^{(4)}_4$ is now given by
\beq
I[box]=R R'\int {d^2 l\over |(l +k+q)|^2 |l|^2 |(l+k+k')|^2 |(l+k' +q)|^2}.
\label{box1}
\eeq
We partial fraction the denominator by writing
\beqa
&& A={1\over l(l + k' +k)}
={1\over (k +k')}\left( {1\over l} -{1\over l + k + k'}
\right)\nonumber \\
&& B={1\over (l+k +q)(l + k' +q)}={1\over (k-k')}
\left({1\over l +k+q}-{1\over l + k' +q}\right),\nonumber
\eeqa
so that
\beqa
&& I[box]= R R'\int {dl dl^*\over 4 |q|^2 |l +\eta|^2}
\left( |A|^2 + |B|^2 + A^* B + A B^*\right)
\label{cbox}
\eeqa
where
\beq
\eta\equiv {(k+q)(k'+q)\over 2 q}
\eeq
In the limit $q\to 0$ one can show that the ``mixed products" $A^*B$ and $AB^*$
give directly that part of the box which we have identified
above as ${\cal K}_2$. As we have discussed previously and will discuss
further below, there are good reasons to think this part of the scale invariant
kernel is a new contribution at NLO which is not related to renormalization
effects.

The partial fractioning technique that is the basis of our analysis introduces
spurious singularities and we need to introduce regulators in order to give
meaning to the complex two-dimensional integrals involved. We
will define $\int_{\Lambda_1}$, $\int_{\mu_1}$ and $\int_{\Lambda_1,\mu_1}$
to be suitable UV, IR and UV-IR regularizations of the corresponding
integrals by {\it defining}
\beqa
&& \int_{\Lambda_1} {dl dl^*\over l(l +\eta)^*}= 2\pi log\,\,
{\Lambda_1\over |\eta|}
\nonumber \\
&& \int_{\Lambda_1\mu_1} {dl dl^*\over |l+\eta|^2}=2 \pi log\,\,{\Lambda_1
\over
\mu_1}. \nonumber \\
\label{tad}
\eeqa
The second integral is discussed further in Appendix A. It is easy to show
that all the spurious UV singularities
introduced by the complex decomposition cancel. The infrared singularities,
instead, in single integrals which are IR divergent, do not cancel.
The analysis of their cancellation is the non trivial part of our analysis.
With the above definitions we obtain e.g.
\beq
I_1 = \int_{\mu_1}{dl dl^*\over |l+ \eta|^2 l (l + k' +q)^*}=
{2 \pi\over \eta (k' + q -\eta)^*}log \,{|k'+q| \mu_1\over |k' +q-\eta||\eta|}
\eeq
and
\beq
\int_{\mu_1}{dl dl^*\over |l + \eta|^2 (l + k + k')(l + k +q)^*}= {2
\pi\over {\eta'}^*(k'-q-\eta')}log{|k'-q|\mu_1\over
|\eta'||k'-q-\eta|},\,\,\,
\eeq
where $\eta'\equiv \eta-k-q $.

We can now evaluate the integrals involving
$A$ and $B$ as follows. The ``mixed" terms give
\beqa
&& I[ A B^*] + c.c.=|k^2-q^2|^2|k'^2-q^2|^2
\int {dl dl^*\,\, A B^*\over |2 q l + (k+q)(k'+q)|^2}
\nonumber \\
&& ={|k^2-q^2|^2|k'^2-q^2|^2\over 4 |q|^2 (k+k')(k-k')^*}
\left({2 \pi\over \eta (k+q-\eta)^*}log{|k+q|\mu_1\over |\eta||k+q-\eta|}
\right.\nonumber \\
&&\left. -{2 \pi\over \eta (k'+q-\eta)^*}log {|k'+q|\mu_1\over
|\eta||k'+q-\eta|} - {2\pi\over (\eta-k-q)^*(k+k'-\eta)}log{|k'-q|\mu_1\over
|\eta-k-q||k+k'-\eta|}
\right.\nonumber \\
&&\left. +{2\pi\over (\eta-k'-q)^*(k+k'-\eta)}log{|k-q|\mu_1\over
|\eta-k'-q||k+k'-\eta|} \right)\,\,+\,\,\,\, c.\,\,c.\nonumber \\
\eeqa
Similarly we obtain
\beqa
&& I[|A|^2]=|k^2-q^2|^2|k'^2-q^2|^2
\int {dl dl^*\,\, |A|^2\over |2 q l + (k+q)(k'+ q)|^2}\nonumber \\
&&={|k^2-q^2|^2|k'^2-q^2|^2\over 4 |q|^2|k+k'|^2}
\left(-{2\pi\over \eta(k+k'-\eta)^*}
log{|k+k'|\mu_1\over \eta (k+k'-\eta)}-{2\pi\over \eta^*(k+k'-\eta)}
\right.\nonumber \\
&&\left.\times log {|k+k'|\mu_1\over |\eta||k+k'-\eta|} +
{4\pi\over |\eta|^2}log{|\eta|\over \mu_1} +{4\pi\over |k+k'-\eta|^2}
log{|k+k'-\eta|\over \mu_1}\right)\nonumber \\
\eeqa
and
\beqa
&&I[|B|^2]=|k^2-q^2|^2|k'^2-q^2|^2
\int {dl dl^*\,\, |B|^2\over |2 q l + (k+q)(k'+ q)|^2}\nonumber \\
&&={2|k^2-q^2|^2|k'^2-q^2|^2\over 4 |q|^2|k-k'|^2}
\left( {2\pi\over |k+q-\eta|^2}log{|k+q-\eta|\over \mu_1}\right.\nonumber \\
&&\left. +{2\pi\over |k'+q-\eta|^2}log{|k'+q-\eta|\over \mu_1}
-{\pi\over (k+q-\eta)^*(k'+q-\eta)}log{|k'+q-\eta||k+q-\eta|\over \mu_1
|k'-k|}\right.\nonumber \\
&&\left. -{\pi\over (k+q-\eta)^*(k'+q-\eta)}log{|k'+q-\eta||k+q-\eta|
\over \mu_1 |k'-k|}\right.\nonumber \\
\eeqa
Moving on to the other connected components of $K^{(4)}_{2,2}$, we obtain
\beqa
&& K_2^{(4)}(k,k',q)=-\left({4\pi|k+q|^2|k'+q|^2\over |k+k'|^2}
log{|k'-q|^2\over \mu_1}
+{4\pi|-k+q|^2|k'+q|^2\over |-k+k'|^2}log{|k'-q|^2\over \mu_1}
\right.\nonumber \\
&&\left.+ {4\pi|k+q|^2|-k'+q|^2\over |k-k'|^2}
log{|k'-q|^2\over \mu_1}
+ {4\pi|-k+q|^2|-k'+q|^2\over |k+k'|^2}log{|k'+q|^2\over \mu_1}\right.
\nonumber \\
&&\left. +{4\pi|k-q|^2|k'-q|^2\over |k+k'|^2}
log{|k +q|^2\over \mu_1}+
{4\pi|k+q|^2|k'-q|^2\over |-k+k'|^2}log{|-k +q|^2\over \mu_1}
\right.\nonumber \\
&&\left. +{4\pi|k-q|^2|k'+q|^2\over |k-k'|^2}
log{|k +q|^2\over \mu_1}+
{4\pi|k+q|^2|k'+q|^2\over |k+k'|^2}
log{|k -q|^2\over \mu_1}\right)\nonumber \\
\eeqa
and
\beqa
&& K_3^{(4)}(k,k',q)=\left({4\pi|k+q|^2|k'+q|^2\over |k+k'|^2}
log{|k +k'|^2\over \mu_1}+
{4\pi|-k+q|^2|k'+q|^2\over |-k+k'|^2}
log{|-k +k'|^2\over \mu_1}\right. \nonumber \\
&&\left. +{4\pi|k+q|^2|-k'+q|^2\over |k-k'|^2}log{|k -k'|^2\over \mu_1}
+{4\pi|-k+q|^2|-k'+q|^2\over |k+k'|^2}
log{|k +k'|^2\over \mu_1}\right)\nonumber \\
\eeqa

\mainhead{4. INFRARED CANCELLATIONS}

In order to prove that the complete kernel, $K^{(4)}_{2,2}$, is IR safe,
and to simplify the notation, let's define ${\cal R}_{\mu_1}$ as
the operation which isolates the infrared sensitive logarithms of all the
components i.e. the coefficient of $log{\mu}_1$. We get (omitting an overall
factor of $2\pi$)
\beqa
{\cal R}_{\mu_1}*I[A B^*]=-{R R'\over 4 |q|^2}
\left({1\over (k+q-\eta)(k'+q-\eta)^*\eta^*(k+k'-\eta)^*} + c.c.\right)
\nonumber\\
\eeqa
leading to
\beq
{\cal R}_{\mu_1}*I[A B^* + A^*B]=8 |q|^2,
\label{nonfo}
\eeq
This shows that the "mixed" contributions are separately IR safe only in the
forward direction i.e. $q=0$.

We similarly obtain
\beq
{\cal R}_{\mu_1}*I[|A|^2]=-4 |q|^2 -{|k-q|^2|k'-q|^2 + |k+q|^2|k'+q|^2\over
|k+k'|^2}
\label{nonfo1}
\eeq
and
\beq
{\cal R}_{\mu_1}*I[|B|^2]=-4 |q|^2 -{|k+q|^2|k'-q|^2 + |k-q|^2|k'+q|^2\over
|k+k'|^2}
\label{nonfo2}
\eeq
Combining these last results with (\ref{nonfo}) we see that
$R_{\mu_1}*I[box]$ is non zero. We conclude that the box diagram is not
separately IR safe.

We also obtain
\beqa
&& {\cal R}_{\mu_1}*{K_4}^{(4)}=-{2\over |k+k'|^2}
\left(|k-q|^2|k'-q|^2+|k+q|^2|k'+q|^2 \right)\nonumber \\
&&-{2\over |k-k'|^2}\left(|k-q|^2 |k'+q|^2 +|k+q|^2 |k'-q|^2 \right)
\nonumber \\
\label{c1}
\eeqa
and, after a quite involved pattern of cancellations,
\beqa
&&{\cal R}_{\mu_1}*(K_2^{(4)} +K_3^{(4)})=
+{2\over |k+k'|^2}( |k'+q|^2 |k-q|^2 +|k'-q|^2 |k+q|^2 )\nonumber \\
&&+{2\over |k-k'|^2}(|k'+q|^2 |k-q|^2 +|k'-q|^2 |k+q|^2)
\label{c2}
\eeqa
Combining (\ref{nonfo}), (\ref{nonfo1}), (\ref{nonfo}), (\ref{c1}) and
(\ref{c2}) we obtain
\beq
R_{\mu_1}*(K^{(4)}_2 + K^{(4)}_3 + K^{(4)}_4) = 0
\label{rmu}
\eeq
showing that the infra-red divergences cancel.

\mainhead{5. SEPARATION OF THRESHOLDS}

The holomorphic factorization properties of ${\cal K}_2$ clearly suggest
that there should be a conformally invariant extension to the non-forward
direction. Since we expect to identify this extension as a new partial-wave
reggeon amplitude we look for a separately infra-red finite
component of the non-forward kernel which satisfies the Ward identity
constraint that it vanish when any $k_i \to 0, ~i=1,..,4$. From the
unitarity analysis of \cite{cw} and the discussion in \cite{cw1} we know that
we should try to isolate the thresholds from the box diagram associated with
logarithms of type 1) discussed in Section 2.

The logarithms  we are interested in are again present in the mixed terms
$AB^*$ and $A^*B$ discussed in the last Section. However, there are also
additional logarithms of the form $q^2 log 4 q^2$, which are associated with
infra-red divergences that appear. If we extract these logarithms we obtain
\beqa
{\cal I}_{AB}(q) &&=-(1-R_q-R_{\mu_1})* I[A B^* + c.c.] \nonumber \\
&&={ 2 \pi (k+q)^2 (k-q)^2(k'+q)^2 (k'-q)^2\over
(k+ k')^2 (k-k')^2}\nonumber \\
&&\times\left( {(k'^2-q^2)(k^2-k'^2) + [qk'][k'k]\over (k+q)^2(k'-q)^2(k'+q)^2}
log (k'+q)^2(k'-q)^2(k+q)^2\right.\nonumber \\
&&\left. -{(k^2-q^2)(k^2-k'^2)+[qk][k'k]\over (k' +q)^2(k-q)^2(k+q)^2}log
(k+q)^2(k-q)^2
(k'+q)^2\right. \nonumber \\
&&\left.- {(k^2-q^2)(k^2-k'^2) +[kq][k'k]\over
(k +q)^2(k'-q)^2(k-q)^2}log (k+q)^2(k-q)^2
(k'-q)^2\right. \nonumber \\
&&\left. +{(k'^2-q^2)(k^2-k'^2) +[k'q][k'k]
\over (k' +q)^2(k-q)^2(k' -q)^2}log (k'+q)^2(k'-q)^2
(k-q)^2 \right). \nonumber \\
\label{ch3}
\eeqa
where we have defined $[qk]\equiv q^*\,k-q\, k^*$. Introducing vectors
$\hat{k}=(k_1,-k_0)$, dual to
$k=(k_0,k_1)$, with the properties $\hat{k}^2=k^2$ and $\hat{k}\cdot k=0$,
\beq
[qk]= 2 i \hat{k}\cdot q.
\eeq

It is straightforward to check that
$$
\eqalign{ {\cal I}_{AB}(q)
&~~ \centerunder{$\longrightarrow$}{\raisebox{-3mm}
{$ q^2 \to 0$ } }
{(k^2-k'^2) k^2 k'^2\over (k+k')^2(k-k')^2} log {k^2\over
k'^2}\cr
&~~= {\cal K}_2 }
\label{scale}
\auto
$$
However, ${\cal I}_{AB}(q) $ has several problems if we wish to identify
it as a non-forward extension of ${\cal K}_2$. It is not infra-red finite in
the sense that the arguments of the logarithms are not ratios of momentum
factors. In addition the behaviour at the thresholds
i.e. at $q \pm k \to 0$ and $q \pm k' \to 0$ is sufficiently singular that
{\it the Ward identities are not satisfied}. That is ${\cal I}_{AB}(q)$ does
not vanish in these limits.

We conclude that the transverse momentum integral corresponding to the
non-forward box diagram does not contain the extension of ${\cal K}_2$ that
we are seeking. Given the limitations of the transverse momentum integral
formalism away from $q^2 = 0$ that we have discussed in \cite{cw} this is,
perhaps, not surprising.

For completeness we also give here the explicit expression for the remainder
of the $O(g^4)$ connected part of $K^{(4)}_{2,2}$. That is if we write
$$
K^{(4)}_2 + K^{(4)}_3 + K^{(4)}_4 = {\cal R}(q) + {\cal I}_{AB}(q)
\auto
$$
then
\beqa
&& {\cal R}(q)={2\pi (k+q)^2(k-q)^2(k'+q)^2(k'-q)^2\over (k+k')^2}\nonumber \\
&& \left( {(k'^2-q^2)(k^2-q^2) + 2 [qk][qk']\over
(k+q)^2(k-q)^2(k'+q)^2(k'-q)^2} log{(k+k')^2\over
(k+q)^2(k'+q)^2(k-q)^2(k'-q)^2} \right. \nonumber \\
&&\left. +{1\over (k+q)^2(k'+q)^2}log (k+q)^2(k'+q)^2 +
{1\over (k-q)^2(k'-q)^2}log(k-q)^2(k'-q)^2\right)\nonumber \\
&& +{ 2\pi (k+q)^2(k-q)^2(k'+q)^2(k'-q)^2\over (k-k')^2}\nonumber \\
&&\times\left({1\over (k+q)^2(k'-q)^2}log (k+q)^2(k'-q)^2 +
{1\over (k'+q)^2(k-q)^2}log (k'+q)^2(k-q)^2 \right.\nonumber \\
&&\left. -{(k'^2-q^2)(k^2-q^2) -2(qk')(qk)\over (k+q)^2(k-q)^2(k'+q)^2(k'-q)^2}
log {(k'+q)^2(k-q)^2(k+q)^2(k'-q)^2\over (k-k')^2}\right)\nonumber \\
&& + {2\pi \over (k+k')^2}log(k+k')^2 (k+q)^2(k'+q)^2
+{2 \pi\over (k+k')^2}log(k'-q)^2(k+q)^2(k'+q)^2\nonumber \\
&& +{2 \pi\over (k+k')^2}log(k+q)^2(k'-q)^2(k-q)^2.\nonumber \\
\label{chi4}
\eeqa

\mainhead{6. THE NON-FORWARD EXTENSION OF ${\cal K}_2$ }

{}From the discussion of the last Section, it is clear that to find
an extension of ${\cal K}_2$ that satisfies the Ward identity constraints,
we must weaken the thresholds in
${\cal I}_{AB}(q) $ at $q \pm k \to 0$ and $q \pm k' \to 0$. A simple way to
achieve this is to remove the denominator in (\ref{cbox}). To retain the
correct dimension we must modify the ``vertex function'' and reduce the
degree of the zeroes at $q \pm k \to 0$, $q \pm k' \to 0$. Consequently we
now define
\beq
{\cal K}_2(k,k',q) =  (q^2 - k^2)(q^2 - k'^2) \int dl dl^* [A^* B + A B^* ]
\eeq
Using extensively the first integral in (\ref{tad}) we obtain
$$
{\cal K}_2(k,k',q) =
\Biggl( { (k^2 - k'^2)(q^2 - k^2)(q^2 - k'^2) \over (k + k')^2 (k - k')^2}
\Biggr) log \Biggl[{(q+k)^2(q-k)^2 \over (q+k')^2(q-k')^2} \Biggr]
\label{ext}
\auto
$$
Clearly
$$
\eqalign{ {\cal K}_{2}(q,k,k')
{}~~\centerunder{$\longrightarrow$}{\raisebox{-3mm}
{$ q^2 \to 0$ } } ~~{\cal K}_2 }
\auto
$$
It is also manifest that ${\cal K}_2(k,k',q) \to 0$ when $q \pm k \to 0$ or
$q \pm k' \to 0$. Consequently ${\cal K}_2(k,k',q)$
satisfies the Ward identity constraints, has all the right symmetries, and has
singularities only at the desired thresholds.

\mainhead{7. CONCLUSIONS}

It is clearly of considerable interest to study the conformal
properties of ${\cal K}_2(q,k,k')$ in the conjugate impact parameter space.
Given the parallel with the leading-order kernel it is natural to expect
that we will find an analogous conformal invariance property. Indeed we
conjecture that the Ward Identity constraints are the crucial feature that
determine the conformally invariant non-forward extension of a kernel with
a holomorphically factorizable spectrum.

It is interesting that to obtain ${\cal K}_2(k,k',q)$ we had to abandon the
transverse momentum integral formalism and go to the complex momentum
formalism of Kirschner. This is clearly related to the natural connection
between the complex momenta formalism and conformal symmetry. It is also
consistent with the limitations of the transverse momentum integral
formalism uncovered in the unitarity analysis of \cite{cw}.

\newpage

\centerline{\bf Acknowledgements}
R. P. thanks the Theory Group at Argonne for its hospitality.
C.C. warmly thanks the Theory Group at the Physics Dept. of the
Univ. of Lecce, Italy, for their hospitality at various stages of this work.

\renewcommand{\theequation}{A.\arabic{equation}}
\setcounter{equation}{0}
\vskip 1cm \noindent
\noindent {\large\bf Appendix A. Regularization of Integrals}
\vskip 3mm

This appendix illustrates in more detail the procedure adopted in the
regularization of the various integrals we encounter. As an example let's
consider
\beq
I_1=\int_{\Lambda_1} {d^2l\over l(l+\eta)^*}
\eeq
in which the complex integration region is defined for
$|l|<\Lambda_1$, since there is an UV divergence.
We rewrite it as a contour integral on the unit circle

\beq
I_1=\int_{0}^{\Lambda_1}d|l|\oint {dw\over i w(1+ w{\eta^*\over |l|})}
\eeq
and perform the contour integral to get

\beq
I_1=2\pi\int_{|\eta|}^{\Lambda_1}d|l|=2\pi log{\Lambda_1\over |\eta|}
\eeq
Complex changes of variables are also allowed
\beqa
&& k\to l+\eta=l'\nonumber \\
&& l^*\to l^* +\eta^*={l'}^*\nonumber \\
&& dl dl^*=dl'd{l'}^*
\eeqa
giving
\beqa
&&\int_{\Lambda_1\mu_1}
{dl dl^*\over |l +\eta|^2}=2\pi
\int_{\mu_1}^{\Lambda_1} {dl' d{l'}^*\over |l'|^2}\nonumber\\
&&=2\pi log{\Lambda_1\over \mu_1}.
\label{set}
\eeqa
Notice that this integral is a ``massless tadpole'' and, as
we know, in dimensional regularization (DR) it is hard to make sense out of it
both in $2+\epsilon$ and in $2-\epsilon$ dimensions. Therefore, massless
tadpoles, in DR are set to be zero. This is not the case in our analysis and
 eq.~(\ref{set}), therefore, has to be handled with a special care.

In order to further illustrate the last  point let's consider the integral
\beq
b=\int {d^2l\over | l-\eta|^2}={1\over i\eta^*}\int_{0}^
{\Lambda_1}d|l|\oint {dw\over (w-{\eta\over |l|})
( {|l|\over \eta^*}-w)}
\eeq
where we have again rewritten the angular integral in a contour form.
The radial integral is ill defined. In our case we get
\beq
b=2\pi\int_{|\eta|}^{\infty}{d|l| |l|\over |l|^2-|\eta|^2} -
4\pi\int_{0}^{|\eta|}{d|l||l|\over |l|^2-|\eta|^2}
\eeq
A careful evaluation then gives
\beqa
&& b=\pi log{4({\Lambda_1}^2-{|\eta|}^2)\over {\mu_1}^2}\nonumber \\
&&= 2 \pi log{\Lambda_1\over \mu_1} +(2\pi log\,\,2).
\eeqa
Notice that this last term $(2\pi log\,\,2)$ has to be omitted in order
to obtain consistent results.
By doing so the result is consistent with
dimensional regularization.

The method, therefore, simply consists of a combination of partial fractioning
of complex propagators with an application of eq.~(\ref{tad}) to evaluate all
the integrals involved.
Partial fractioning introduces spurious singularities at intermediate stages,
which cancel only if the integral is well defined.
For example, let's consider

\beqa
&& \int_{\mu_1} {dl dl^*\over |l(l+ k +k')|^2}=
\int {dl dl^*\over |k+k'|^2}\left|{1\over l} -{1\over l + k +k'}\right|^2
\nonumber \\
&&= {2\pi\over |k+k'|^2}\left( log{\Lambda_1\over \mu_1}-
log{\Lambda_1\over |k+k'|}\right)\nonumber \\
&&={2 \pi\over |k+k'|^2} log {|k+k'|\over \mu_1},
\label{check}
\eeqa
where the $Log\Lambda_1$ terms cancel in the final result.
It can shown that all such terms disappear in the final expression of the box,
as expected.
Notice that
(\ref{check}) is exactly what one expects from dimensional regularization
after expanding the result for the self energy diagram in $D=2 +\epsilon$
dimensions and introducing a renormalization scale $\mu_1$.
In fact
\beq
\int {d^{2+\epsilon}\over l^2 (l + k + k')^2}= {\pi\over (k+k')^2}
Log \, {(k+k')^2\over {\mu_1}^2 } + \,\,\,\,\, O(\epsilon).
\eeq

The derivation of $I_1$, given in section 3, proceeds as follows.
After a partial fractioning we get

\beq
I_1={1\over \eta (k+q-\eta)^*}\int dl dl^*\left(
{1\over l(l+\eta)^*}- {1\over l(l+k+q)^*} - {1\over |k+\eta|^2} +
{1\over (l+\eta)(l+k+q)^*}\right)
\eeq
Using (\ref{tad}) we get
\beq
I_1={2\pi\over \eta(k+q-\eta)^*}\left( log{\Lambda_1\over |\eta|} -
log{\Lambda_1\over |k+q|} -log{\Lambda_1\over \mu_1}
 +log{\Lambda_1\over |k+q-\eta|} \right).
\eeq
Combining all the 4 terms together we get the result given in section 3.

As another example we consider
\beq
I_2 =\int {d^2l\over |l+\eta|^2(l + k +q)^*(l +k'+q)}
\eeq
After partial fractioning the denominator
\beqa
&&{1\over (l+\eta)(l+\eta)^*(l+k+q)^*(l+k'+q)}=
{1\over (k+q-\eta)(k'+q-\eta)^*}\nonumber \\
&& \left({1\over |l +\eta|^2}
-{1\over (l+\eta)(l+k'+q)^*}\right.\nonumber \\
&&\left. -{1\over l+k+q)(l+\eta)^*}+{1\over (l+k+q)((l+k'+q)}\right)
\eeqa
and closing contour integrations in the various sub-integrals we get
\beqa
&&I_2 ={2 \pi\over (k+q-\eta)(k'+q-\eta)^*}
\left(log{\Lambda\over \mu_1} - {1\over |k'+q-\eta|}log{\Lambda\over
|k'+q-\eta|}\right.\nonumber \\
&&\left. -{1\over |k+q-\eta|}log{\Lambda\over |k+q-\eta|}+
{1\over |k'-k|}log{\Lambda\over |k'-k|}\right)\nonumber \\
&&={2\pi\over(k+q-\eta)(k'+q-\eta)^*}log{|k+q||k-q||k'+q||k'-q|\over
4 |q|^2\lambda |k-k'|}
\eeqa

An additional check on the consistency of the methods of regularization,
after partial fractioning, comes from the cancellation of the $q^2 log\,q^2$
terms in the non-forward kernel. We have briefly mentioned this important
point in Section 5. Since the proof is not obvious, we briefly sketch
it here. We introduce a new subtraction, denoted as $R_q$, which isolates the
$log \,\, 4 q^2$ terms in the ``mixed'' contributions. After some appropriate
manipulations we get
\beqa
 R_qI[A B^* + c.c.]=- {R R'\left[(k+q-\eta)^*(k'+q-\eta)\eta(k+k'-\eta) + c.c.
\right]\over 4 |q|^2|k +q-\eta|^2 |k'+q-\eta|^2|\eta|^2|k+k'-\eta|^2}
= 8 |q|^2, \nonumber \\
\eeqa
(an overall factor of $2 \pi$ has been omitted).
Similarly
\beqa
&& R_q*I[|A|^2]= {R R'\over 4 |q|^2 |k+k'|^2}
\left(-{1\over \eta (k + k'-\eta)^*}- {1\over \eta^* (k + k'-\eta)}
\right.\nonumber \\
&&\left.\,\,\,\,\,\,\,\,\,\, -{1\over |\eta|^2}
-{1\over |k+k'-\eta|^2}\right)  \,\,\, =-4 |q|^2
\eeqa
and
\beqa
&& R_q*I[|B|^2]= {R R'\over 4 |q|^2 |k-k'|^2}
\left(-{1\over |k+q-\eta|^2}-{1\over |k'+q-\eta|^2}\right.\nonumber \\
&&\left.\,\,\,\,\,\,\,\,\,\, +{1\over (k+q-\eta)(k'+q-\eta)^*} +
{1\over (k+q-\eta)(k'+q-\eta)}\right) \,\,  =-4 |q|^2.
\eeqa
Therefore we have the identity
\beq
R_q*I[box]=R_q\left(I[A B^* +c.c.] + I[|A|^2] + I[|B|^2]\right)=0
\label{rqu}
\eeq
as should be the case.

The expression for $(1-R_q -R_{\mu_1})*I[A B^* + c.c.]$ has been given in
Section 5. Using this, together with the identity
\beq
I[box]=(1- R_q)*I[A B^* + c.c.] +
(1-R_q)*(I[|A^2|] + I[|B|^2]).
\label{rr}
\eeq
we obtain for the other parts of the box
\beqa
&& (1-R_q-R_{\mu_1})*I[|A|^2]={2 \pi R R'\over |k+k'|^2}\nonumber \\
&&\times\left(
{(|k'|^2- |q|^2)(|k|^2-|q|^2) + [qk][qk']\over R R'}
log{|k+k'|^2\over R R'}\right. \nonumber \\
&&\left.  +{2 \over |k+q|^2|k'+q|^2}log|k+q|^2|k'+q|^2
+{2\over |k-q|^2|k'-q|^2}log|k-q|^2|k'-q|^2\right), \nonumber \\
\eeqa
and
\beqa
&& (1-R_q-R_{\mu_1})*I[|B|^2]={2 \pi R R'\over |k -k'|^2}\nonumber \\
&&\times\left(
{(|k'|^2- |q|^2)(|k|^2-|q|^2) - [qk][qk']\over R R'}
log{|k-k'|^2\over R R'}\right. \nonumber \\
&&\left.  +{2 \over |k+q|^2|k'-q|^2}log|k+q|^2|k'-q|^2
+{2\over |k-q|^2|k'+q|^2}log|k-q|^2|k'+q|^2\right).\nonumber \\
\eeqa

Using these relations and the identity (\ref{rmu}) we obtain an expression
for the full NLO connected kernel of the form
\beqa
&& K^{(4)}_2 +K^{(4)}_3 +K^{(4)}_4= K^{(4)}_2 +K^{(4)}_3 + I[box]
\nonumber \\
&& =(1-R_q-R_{\mu_1})*(K^{(4)}_2 +K^{(4)}_3) +
(1-R_q-R_{\mu_1})* I[A B^* + \nonumber \\
&& +(1-R_q-R_{\mu_1})*(I[|A|^2] + I[|B|^2]).
\eeqa

\renewcommand{\theequation}{B.\arabic{equation}}
\setcounter{equation}{0}
\vskip 1cm \noindent
\noindent {\large\bf Appendix B. Spectrum Evaluation. }
\vskip 3mm

This appendix contains some comparison of  two possible ways to evaluate the
spectrum of the various kernel components that we have discussed. The first
method, already presented in \cite{cw1}, is based on the use of dimensional
regularization. The second method has been briefly discussed by
Kirschner\cite{rk}. In the case of the forward kernel, the evaluation of the
spectrum of the new partial wave component is more easily performed by
 using dimensional regularization. However, we
anticipate that the second approach may be useful for studying the
non-forward kernel and so we give more details here.

A simple treatment of the bubble diagram or 2-point function,
$J_1(k)$, illustrates both methods.
In the approach of \cite{rk} we work in $D=2$ and regulate the 2-point
function by a cutoff ($\lambda$) using the integral representation
\beq
\theta[|k'-k|-\lambda]={1\over 2 \pi i}
\int_{-i\infty +\delta}^{i\infty +\delta}{d\omega\over \omega}
\left[{|k'-k|^2\over \lambda^2}\right]^\omega.
\label{theta}
\eeq
The contour in $\omega$ is closed on the right half plane or on the
left half plane depending on whether $|k'-k|<\lambda^2$ or $|k'-k|>\lambda^2$
respectively. All the diagrams will now depend on $\omega$ and the
final result in the calculation of the spectrum is obtained by extracting the
residui for $\omega=0$ of the eigenvalues (i.e. performing the integral over
$\omega$ at the end).

This method allows us to work directly at $D=2$ without the need of a mass
cutoff in the propagators.
Therefore, we regulate $J_1(k)$ by
\beq
J_{1,reg}(k)={1\over (2 \pi i)^2}
\int {d\omega d\omega_0\over \omega \omega_0 }
\int {d^2 k'\over |k'|^2 |k-k'|^2}
\theta[|k'-k|-\lambda]\theta[|k'|-\lambda].
\eeq
(For simplicity we omit an overall factor $1/(16 \pi)^3$ compared to (\ref{j1})
).
After inserting the representation of the step-function given by
(\ref{theta}) and after performing the integral over the loop momentum
with the help of the formula (with $D=2$)
\beqa
&& I[R,S]=\int {d^D k\over [(k-q)^2]^R[k^2]^S}\nonumber \\
&&={\pi^{D/2}\Gamma[D/2-R]\Gamma[D/2-S]\Gamma[R +S-D/2]\over
\Gamma[R]\Gamma[S]\Gamma[D-R-S][q^2]^{R+S-D/2}},
\eeqa
it is straightforward to obtain
\beqa
&&J_{1,reg}(k)={1\over (2\pi i)^2 k^2}
\int {d\omega d\omega_0\over \omega \omega_0}
\left({k^2\over \lambda^2}\right)^{\omega +\omega_0}C(\omega,\omega_0)
\eeqa
where
\beq
C(\omega,\omega_0)=
({1\over \omega}+{1\over \omega_0})f(\omega,\omega_0),
\eeq
with
\beq
f(\omega,\omega_0)=
{\pi\Gamma[1-\omega-\omega_0]\Gamma[1+\omega]\Gamma[1+\omega_0]\over
\Gamma[1-\omega_0]\Gamma[1-\omega]\Gamma[1+\omega+\omega_0]}.
\eeq

After expanding in $\omega$ and $\omega_0$ and picking up the residui at the
single poles in both variables we get
\beqa
&& J_{1,reg}(k)={1\over (2 \pi i)^2 k^2}
\int {d\omega d\omega_0\over \omega \omega_0 }
( 2\pi Log {k^2\over \lambda^2} \,\,+\,\,...)
f(\omega,\omega_0)\nonumber \\
&& = {2 \pi\over k^2} Log{k^2\over \lambda^2},
\eeqa
which is the usual (cutoff) expression.
As shown in \cite{cw1}, the eigenfunctions of the NLO unitarity kernel
are of the form $f_{n,\nu}(k)=(|k|^2)^{1/2+i\nu}({k\over |k|})^n$,
as in the lowest order case. Convoluting $J_1(k)$ with these
eigenfunctions one gets the eigenvalue equation
\beqa
&& J_{1,reg}*f_{n,\nu}\equiv (k^2)^2 \int {d^2\, k'\over |k'|^2 |k|^2}
J_{1 reg}(k-k')(|k'|^2)^{1/2 + i \nu}\left({k'\over k}\right)^n\nonumber \\
&& =\lambda_s(a,b) f_{n,\nu}.
\label{div}
\eeqa

Notice that in (\ref{div}) we have divided by the factor $1/(k^2 k'^2)$
which appear in the definition of the convolution product \cite{cw1}, and
introduced a vertex factor $(k^2)^2$, as discussed in Section 2. The
singularity
at $k'=0$ does not need regularization since it is taken care of by the
power behaviour of the eigenfunctions.
We get
\beqa
 J_{1,reg}*f_{n,\nu}=
{k^2\over (2 \pi i)^2}
\int {d\omega d\omega_0\over \omega \omega_0}{C(\omega,\omega_0)\alpha(k)\over
(\lambda^2)^{\omega +\omega_0}},
\eeqa
where
\beq
\alpha(k, a,b)=\int {dk' d\theta |k'|^n e^{i n \theta}\over
(|k'|^2)^a(|k'-k|^2)^b}.
\eeq
with $a=1/2-i\nu +n/2$ and $b={1 -\omega -\omega_0}$.
The angular integral has branch cuts, due to the $\omega$ terms
in the exponents of the denominator. The same difficulty appears in
dimensional regularization.

In order to understand this last issue we illustrate the point in detail.
The structure of the angular integral in $\alpha(k)$ is of the form
\beqa
&&I=\int_{0}^{2\pi}{ d\theta e^{i n \theta}\over (1-z \,cos\theta)^{\eta}}
\nonumber \\
&&=i\,(-1)^{\eta + 1}2^\eta\oint {d w w^{n-1+\eta}\over (z w^2 -2 w +z)^\eta}
\nonumber \\
&& =i\,\,2^\eta \oint {dw  (-1)^{\eta +1}
w^{n+1-\eta}\over z^\eta (w-w_1)^\eta(w-w_2)^\eta}
\label{lastl}
\eeqa
with cuts between $[0,w_1=1/z(1- \sqrt{1-z^2})]$ and
$[w_2=1/z(1 +\sqrt{1 -z^2}),\infty]$.
We rewrite it in the form
\beq
I= i (-1)^{\eta +1}(e^{2 \pi i \eta}-1)
\int_{0}^{w_1} {d\,w \,w^{n+1-\eta}\over |w_1-w|^\eta |w_2-w|^\eta},
\label{thei}
\eeq
where we have used the expression of the discontinuity of the factor in the
integrand ($g(w)=-i/\left((w_1-w)^\eta (w_2-w)^\eta\right)$) on the first
interval $[0,w_1]$
\beqa
&& g^+(w)-g^-(w)=-i {e^{2 \pi i\eta}-1\over
|w_1-w|^{\eta}|w-w_2|^\eta}\nonumber \\
&&=2 (-1)^\eta {\pi\over B[\eta,1-\eta]|w_1-w|^\eta |w-w_2|^\eta},\nonumber \\
\eeqa
where we have used
$sin\,\pi\,\eta={\pi/ B[\eta,1-\eta]}$ by an analytic continuation,
and $B[x,y]$ is the beta-function.

We now obtain
\beqa
&& I= {2^{\eta+1 \pi}\over B[\eta,1-\eta]}{z^\eta w_1^n\over w_2^\eta}
\int_{0}^{1} {dx  x^{n-1+\eta}\over |x-1|^\eta ((w_1/w_2) x -1)^\eta}
\nonumber \\
&& ={2^{\eta +1}\pi \over z^\eta n B[\eta,n]}{w_1^n\over w_2^\eta }
\,\,F_{2,1}[\eta,n+\eta,1 +n,{w_1\over w_2}].\nonumber \\
\eeqa
having used as a definition of the hypergeometric function
\beq
F_{2,1}[a,b,c,z]={1\over B[b,c-b]}\int_{0}^{1}t^{b-1}(1-t)^{c-b-1}(1-t z)^{-a}
\,\, dt
\eeq

Along the same lines, following the derivation presented above, one can easily
evaluate the integral
\beqa
&& I'[\eta]=\int_{0}^{2\pi} {d\theta \,e^{i n \theta}\over (1+ v^2 -2 v\, cos\,
\theta)^\eta}
\nonumber \\
&& = i\oint {dw (-1)^{\eta +1}w^{n+1-\eta}\over
v^\eta(w-w_1)^\eta(w-w_2)^\eta},
\nonumber \\
\eeqa
where now the cuts are between $[0,w_1=v]$ and $[w_2=1/v,\infty]$.
Notice that (\ref{lastl}) and (\ref{thei}) are quite general. Therefore it
is a simple exercise to show that
\beq
I'[\eta]={2 \pi v^n\over n B[\eta,n]} F_{2,1}[\eta,\eta +n,n+1,v^2].
\eeq
The radial integral in $k$ can now be done and the result, as we are going
to show, can be expressed in terms of simple hypergeometrics or also as a
string of products of Gamma-functions. This second form is obtained by using
the Schwinger parametrization of the integrand.
Notice that in the evaluation of the spectrum of ${\cal K}_2$ these
difficulties are not present \cite{cw1} since the angular integral has just
two single poles - the angular integration being embedded in D-dimensions.

In dimensional regularization we define
\beqa
&& J_{1}*f_{n,\nu}\equiv (k^2)^2 \int {d^D\, k'\over |k'|^2 |k|^2}
J_{1}(k-k')(|k'|^2)^{1/2 + i \nu}\left({k'\over k}\right)^n\nonumber \\
&& =\lambda_s(a,b) f_{n,\nu}.
\label{div1}
\eeqa
with $D=2 +\epsilon$ and embed $\theta$ in a D-dimensional angular space
parameterized by $(\theta_1,\theta_2,...,\theta_{D-1})$ by assuming
$\theta\equiv \theta_{D-1}$.
Using the expression of $I'[\eta]$ given above it is not hard to show that
\beqa
&&\int_{0}^{2\pi} {d\,\theta e^{i n \theta}\over (k^2 + k'^2 -2 k k' cos(\theta
-\chi))^\eta}\nonumber \\
&&= \theta[k'-k] {2 \pi e^{i n \chi}\over (k'^2)^{2-D/2} n B[\eta,n]}
\left({k'\over k}\right)^n F_{2,1}[\eta,\eta +n,n+1,(k/k')^2]\nonumber \\
&& + \theta[k-k'] {2 \pi e^{i n \chi}\over (k^2)^{2-D/2} n B[\eta,n]}
\left({k\over k'}\right)^n F_{2,1}[\eta,\eta +n,n+1,(k'/k)^2],
\nonumber \\
  \eeqa
with $cos\,\chi=k\cdot \hat{x}$ and $cos\,\theta=k'\cdot \hat{x}$ and
$\eta= 2-D/2$.
Then we get
\beq
 J_{1}*f_{n,\nu}={2 \pi^D\Gamma[D/2-1]^2\Gamma[2-D/2+n]\over
\Gamma[n+1]\Gamma[D-2]\Gamma[D/2]}(\sigma_1 +\sigma_2)(k^2)^{D-2} f_{n,\nu},
\eeq
with
\beqa
&&\sigma_1=\int_{0}^{1} dx \,\,x^{n+ (D-2) +2 i
\nu}F_{2,1}[\eta,\eta+n,n+1,x^2]
\nonumber \\
&&\sigma_2=\int_{0}^{1} {dx\over x^{(2 D-4) -n +2 i\nu}}
F[\eta,\eta+n,n+1,x^2].
\eeqa
Notice that the spurious factors containing ${(k^2)}^{D-2}$, with
$D=2+\epsilon$
 are eliminated at the end, when all components of the spectrum are combined
and the singularities cancel. One easily gets
\beqa
&&\sigma_1={1\over \rho_1}F_{3,2}[\eta,\eta+1\,|\,\rho_1,n+1,\rho
+1,1]\nonumber \\
&& \sigma_2={1\over \rho_2}F_{3,2}[\eta,\eta+1,\rho_2\,
|\,n+1,\rho_2 +1,1],\nonumber \\
\eeqa
with $\rho_1=n/2-1/2 +D/2 + i\nu$ and $\rho_2=5/2 - D -2 -i \nu + n/2$.

One reason for expecting that the cutoff regularization might turn out to be
useful in the investigation of the spectrum of the nonforward kernel is
that, at least to leading order, the eigenfunctions are given by conformal
partial waves, which are known in $D=2$ \cite{lip}. To our knowledge,
however, a direct calculation of the spectrum using these eigenfunctions
has not been attempted, even for the BFKL kernel.

In order to conclude our illustration of the method of calculation of the
spectrum for the 2-point function, we reconsider $\alpha(k,a,b)$,
which we rewrite in exponential form
\beqa
&&\alpha(k,a,b)=\int_{0}^{1}d w_1 dw_2\delta(1-w_1-w_2)\nonumber \\
&&\times
\int d^2k'\int_{0}^{\infty}dx\,\, x^{a+b-1}
e^{-x(\vec{k'}- \vec{k} w_2)^2 +x \vec{k}^2
{w_2^2 -x w_2 \vec{k}^2 }}
{w_1^{a-1}w_2^{b-1}\over \Gamma[a-1]\Gamma[b-1]}
 k'^n.
\label{comp}
\eeqa
To obtain (\ref{comp}) we have used the
Schwinger parametrization of the propagators and
performed a scaling on the integration parameters by $x$.

Notice that we have used a mixed real and complex notation
(for instance $\vec{k}\cdot \vec{k'}=1/2(k k'^* + k^* k')$, and so on)
for convenience. Notice, in particular, that $k'^n$ is a complex vector.
It is easy to show that
\beq
\int d^2k' k'^n e^{-x (k'+k w_2)^2}={\pi\over x }k^n {w_2}^n.
\label{shift}
\eeq
(\ref{shift}) is easily derived by a complex change of coordinates in the
momentum integration and using the complex expansion $(k+k')^n
=\sum_{p=0}^n n!/(k! (n-k)!) k^p k'^{n-p}$. Only the $n=p$ term is nonvanishing
after angular integration.
The integration over $x$ is also gaussian and we get
\beqa
&&\alpha(k,a,b)={\pi k^n\over (|k|^2)^{a -1}
 \Gamma[a-1]\Gamma[b-1]}\Gamma[a+b-1]
\left( {k^2\over \lambda^2}\right)^{\omega +\omega_0}
\int_{0}^{1} d w_2 (1-w_2)^{-b}w_2^{n-a}\nonumber \\
&&= \lambda_s(a,b) f_{n,\nu},
\eeqa
(since $a-1= -1/2 +n/2 -i\nu$ ).
Notice that the additional factor $(k^2/\lambda^2)^{\omega + \omega_0}$
is removed after the final integration in the $\omega$ variables
which sets $\omega=\omega_0=0$.
After some manipulations
we obtain
\beq
\lambda_s(a,b)={\pi\Gamma[a+b-1]\Gamma[1-a]\Gamma[1-b]B[1+n -a,1-b]}.
\eeq
Finally, after performing the integrals over $\omega, \omega_0$ in the
expression above, only the residui at the simple poles in these variables
survive. A similar approach can be followed also in dimensional regularization.
We conclude by recalling that the observation in \cite{cw1} that part of the
$O(g^4)$ forward kernel is simply the square of the $O(g^2)$
Lipatov kernel allowed us to write down immediately the corresponding spectrum.

\end{document}